\pdfoutput=1
\documentclass[
    ,final            % use final for the camera ready runs
  ]
  {aipproc}

\layoutstyle{6x9}

\begin{document}

\title{Thermal Dileptons from Hot and Dense Strongly Interacting Matter}


\author{Hans J. Specht for the NA60 Collaboration}{
  address={Physikalisches Institut, Universit\"{a}t Heidelberg, Heidelberg, Germany}
}

%%\author{for the NA60 Collaboration}{
%%  address={<common address for author2 and author3>}
%%}

%%\author{<author3>}{
%%  address={<common address for author2 and author3>}
%%  ,altaddress={<author1 address>} % additional visiting address
%%}

\begin{abstract}
 The NA60 experiment at the CERN SPS has studied muon-pair production
 in 158A GeV In-In collisions. The unprecedented precision of the data
 has allowed to isolate a strong excess of pairs above the known
 sources in the whole invariant mass region 0.2$<$M$<$2.6 GeV. The
 (mostly) Planck-like shape of the mass spectra, exponential m$_{T}$
 spectra, zero polarization and the general agreement with
 thermal-model results allow for a consistent interpretation of the
 excess dimuons as thermal radiation from a randomized system. For
 M$<$1 GeV, the process
 $\pi^{+}\pi^{-}\rightarrow\rho\rightarrow\mu^{+}\mu^{-}$
 dominates. The associated space-time averaged $\rho$ spectral
 function shows a nearly diverging width in approaching chiral
 symmetry restoration, but essentially no shift in mass. Some
 in-medium effects are also seen for the $\omega$, but not for the
 $\phi$. For M$>1$ GeV, the average temperature associated with the
 mass spectrum is about 200 MeV, considerably above T$_{c}$=170 MeV,
 implying a transition to dominantly partonic emission sources in this
 region. The transition itself is mirrored by a large jump-like drop
 in the inverse slope of the transverse mass spectra around 1 GeV.
\end{abstract}

\maketitle

%%%%%%%%%%%%%%%%%%%%%%%%%%%%%%%%%%%%%%%%%%%%
%% MAINMATTER
%%%%%%%%%%%%%%%%%%%%%%%%%%%%%%%%%%%%%%%%%%%%

\section{INTRODUCTION}

  At high temperature and baryon densities, QCD predicts strongly
  interacting matter to undergo a phase transition from hadronic
  constituents to a plasma of deconfined quarks and gluons. At the
  same time chiral symmetry, spontaneously broken in the hadronic
  world, is restored. High-energy nuclear collisions provide the only
  way to investigate this issue in the laboratory. Among the different
  observables used in this context, lepton pairs are particularly
  attractive. In contrast to hadrons, they directly probe the entire
  space-time evolution of the expanding fireball formed in such a
  collision, escaping freely without final-state interactions. At low
  invariant masses M$<$1 GeV (LMR), thermal dilepton production is
  largely mediated by the light vector mesons $\rho$, $\omega$ and
  $\phi$. Among these, the broad $\rho$ (770) is by far the most
  important, due to its strong coupling to the $\pi^{+}\pi^{-}$
  channel and its life time of only 1.3 fm, making it subject to
  regeneration in the much longer-lived fireball. It has therefore
  been considered since long as the prime probe for ``in-medium
  modifications'' of hadron properties, including even the QCD phase
  boundary itself~[1-5]. At intermediate masses M$>$1 GeV (IMR), where
  hadronic spectral functions become increasingly uniform, Planck-like
  thermal radiation is expected from both hadronic and partonic
  sources. Dileptons with their independent variables mass and
  transverse momentum should be able to unambiguously differentiate
  between the two (unlike real photons).

  Experimentally, excess dileptons above the known sources have been
  found and reported upon before~[6-9], starting around 1995. A short
  review, including also the preceding (very influential) pp era and
  theoretical milestones, has been published in
  \cite{Specht:2007ez}. Unfortunately, insufficient data quality in
  terms of statistics and mass resolution as well as the unclear role
  of non-thermal sources like open charm hindered final clarification
  on all fronts up to recently. A large step forward in technology,
  leading to completely new standards of the data quality in this
  field, has now been achieved by the ``third-generation'' experiment
  NA60 at the CERN SPS. The central results have already been
  published~[11-17]. The present paper shortly reviews them, but
  places also particular emphasis on the quantitative understanding of
  the thermal-radiation part in connection with the medium
  modification of hadrons (one of the main topics of the workshop),
  just to create confidence towards a community which is usually
  exposed to ``elementary'' reactions rather than heavy-ion
  collisions.

\section{MAJOR ANALYSIS STEPS}

The NA60 apparatus \cite{Arnaldi:2008er} combines the muon
spectrometer previously used by NA10/NA38/NA50 with a novel
radiation-hard, silicon-pixel vertex telescope, placed inside a 2.5 T
dipole magnet between the target region and the hadron absorber. The
matching of the muon tracks in the two spectrometers, both in
coordinate and momentum space, greatly improves the dimuon mass
resolution, reduces the combinatorial background from $\pi$ and $K$
decays, and allows to measure the dimuon offset with respect to the
primary interaction vertex to tag open charm decays. The traditional
high luminosity of dimuon in contrast to dielectron experiments is
still conserved.

The data presented in this paper were taken for 158A GeV In-In
collisions. The analysis was done in three major steps: assessment and
subtraction of the background, identification of excess dimuons above
the known sources and their subtraction from the latter, and full
correction for experimental acceptance in all variables. The last two
steps were never done in this field before and are thus unique to
NA60.

The left panel of Fig.1~\cite{Arnaldi:2006jq,Damjanovic:2006bd} shows
the opposite-sign, background and signal dimuon mass spectra,
integrated over all collision centralities (determined by the track
multiplicity in the Si telescope). The combinatorial background has
been determined with an accuracy of about 1\% by using a mixed-event
technique~\cite{Arnaldi:2008er}, the much weaker background of
incorrect (``fake'') matches between the two spectrometers by an
overlay Monte Carlo method. After subtraction of the total background,
the resulting net spectrum contains about 440\,000 events, exceeding
previous results by up to 3 orders of magnitude in effective
statistics, depending on mass.

%%%%%% Fig.1
\begin{figure*}[t]
\includegraphics*[width=0.35\textwidth]{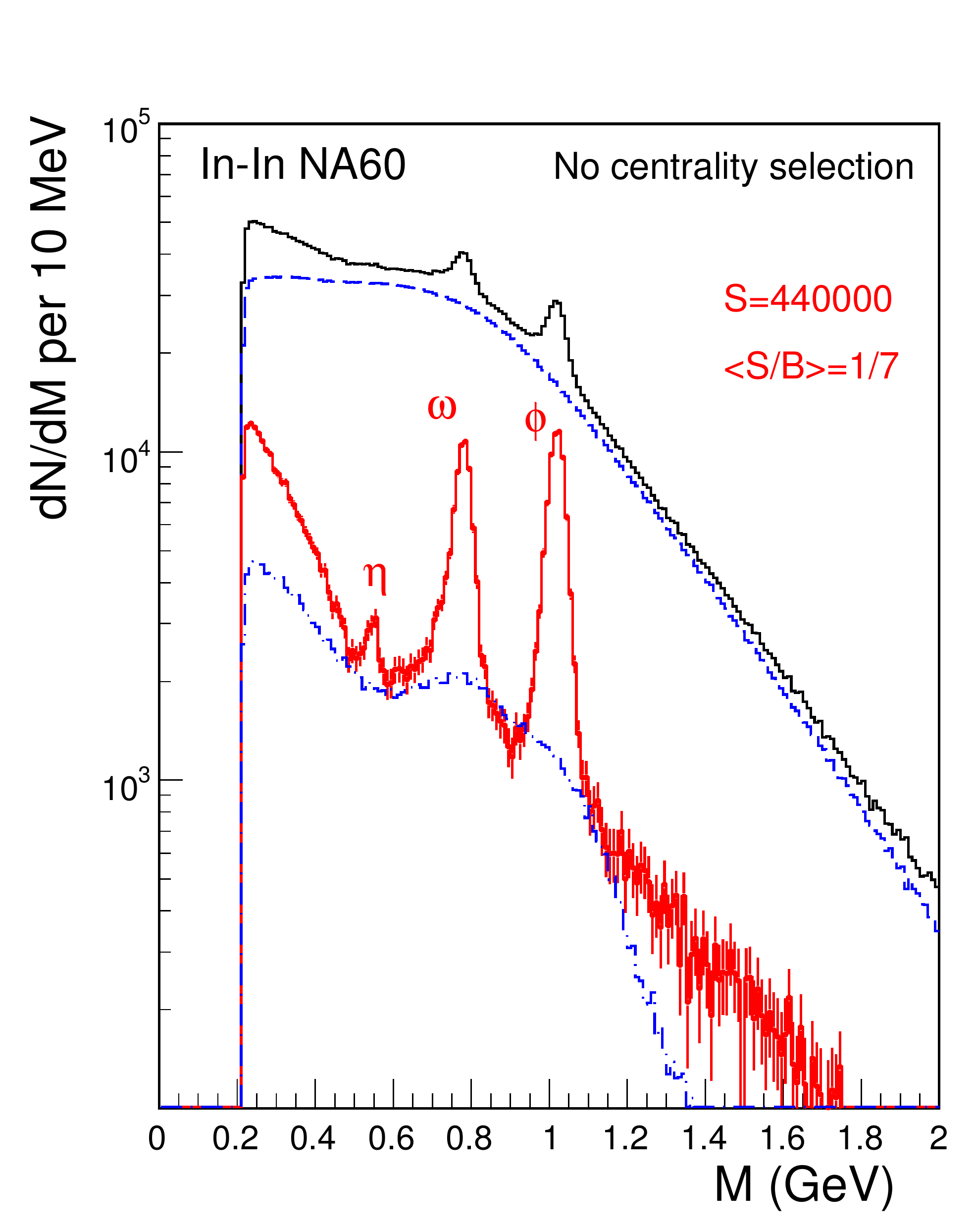}
\includegraphics*[width=0.35\textwidth]{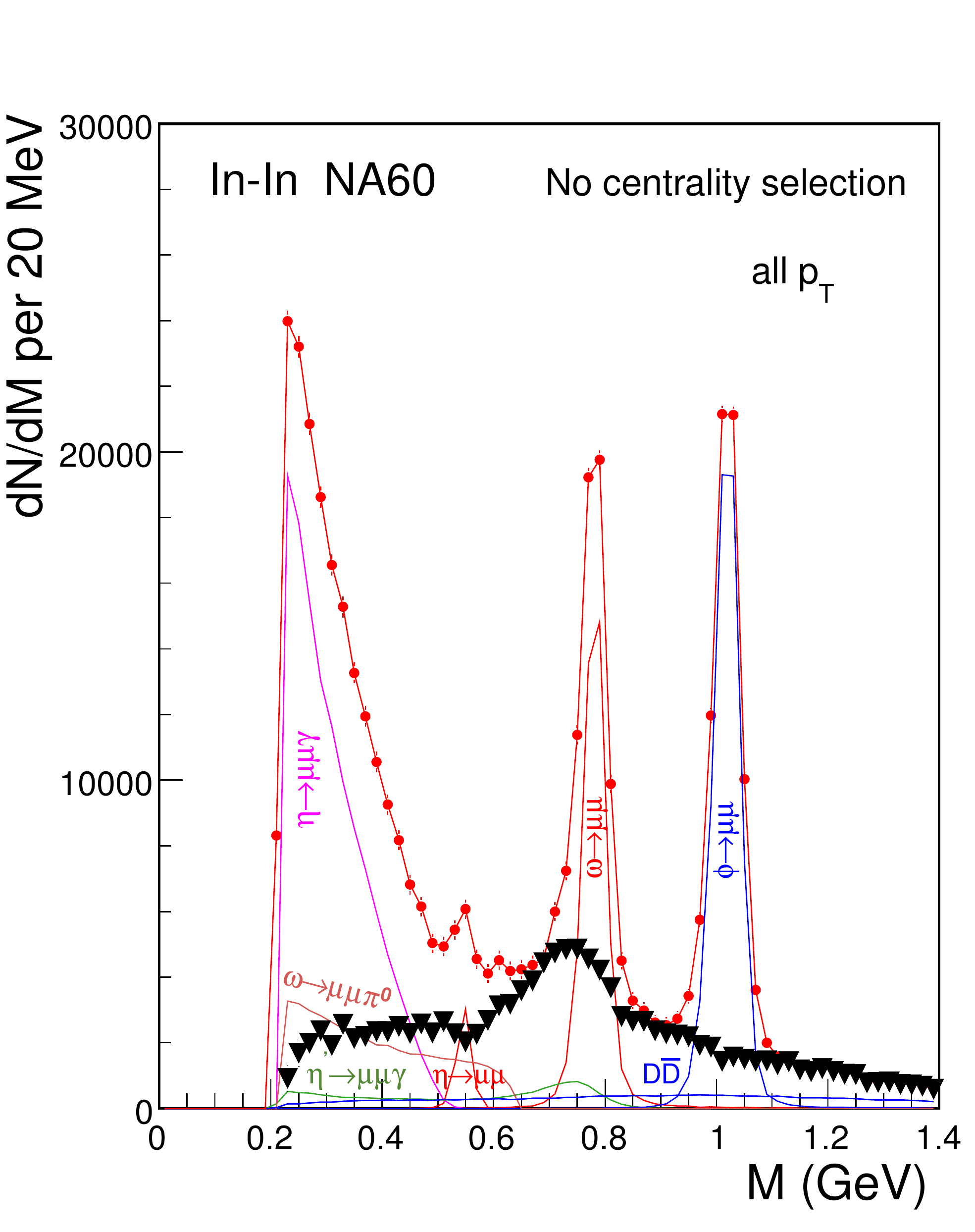}
\includegraphics*[width=0.35\textwidth]{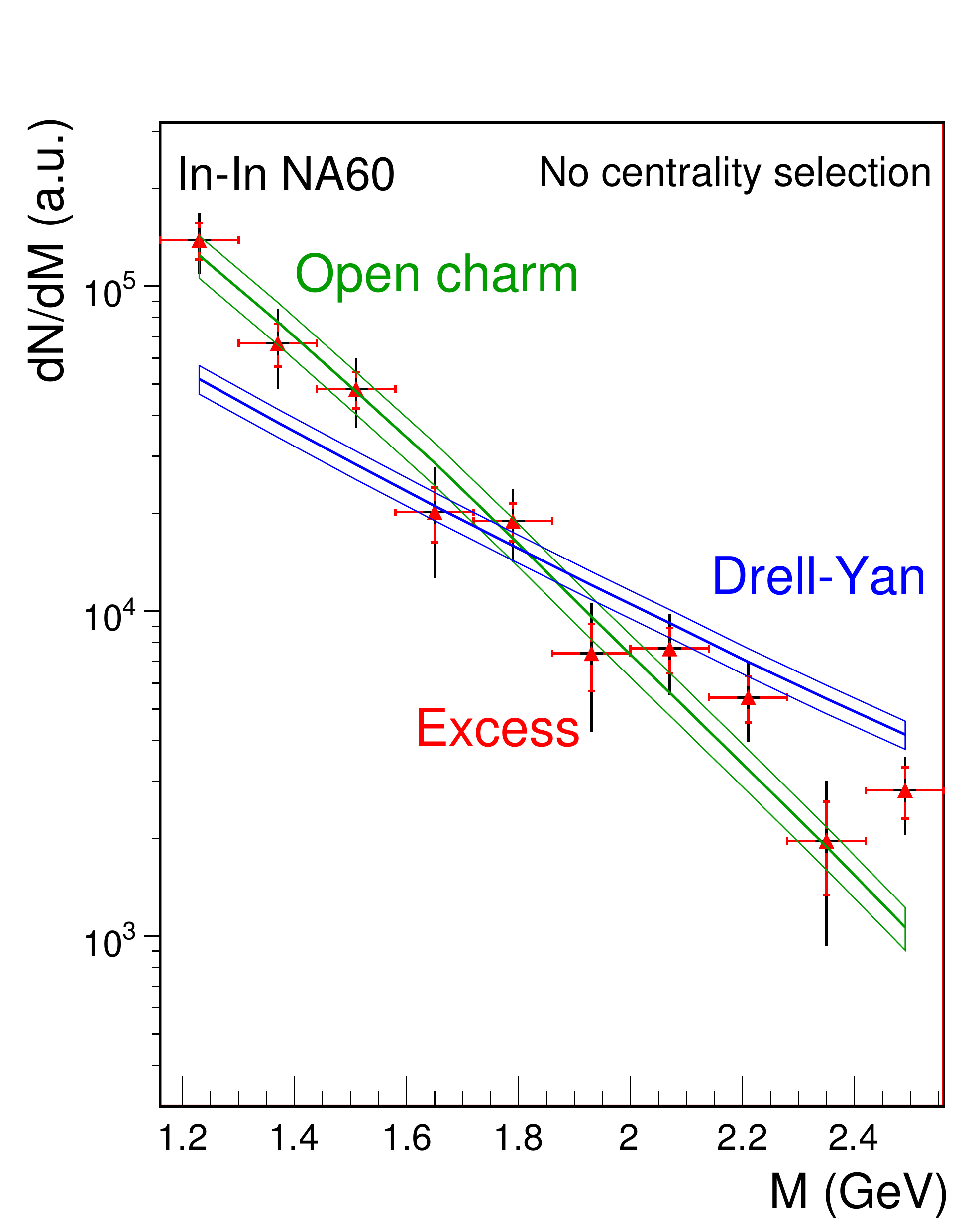}
\caption{Left Panel: Mass spectra of the opposite-sign dimuons (upper
histogram), combinatorial background (dashed), signal fake matches
(dashed-dotted), and resulting signal (histogram with error
bars). Middle Panel: Net mass spectrum before (dots) and after
subtraction (triangles) of the known decay sources. Right panel:
Centrality-integrated mass spectra of Drell-Yan, open charm and the
excess dimuons (triangles), here already acceptance-corrected.}
\label{fig1}
\end{figure*}

The enlarged LMR part of the net spectrum is shown in the middle panel
of Fig.1~[11-14,16]. It is dominated by the known sources: the
electromagnetic two-body decays of the completely resolved $\eta$,
$\omega$ and $\phi$ resonances, and the Dalitz decays of the $\eta$,
$\eta^{'}$ and $\omega$. While the peripheral ``pp-like'' data are
quantitatively described by the sum of a ``cocktail'' of these
contributions together with the $\rho$ and open
charm~\cite{Damjanovic:2006bd}, this is not true for the more
centrally weighted total data shown in Fig.1, due to the existence of
a strong excess of pairs. The high data quality allows to isolate this
excess with a priori {\it unknown characteristics} without any fits:
the cocktail of the decay sources is subtracted from the total using
{\it local} criteria which are solely based on the measured mass
distribution itself. The $\rho$ is not subtracted. The excess
resulting from this difference formation is illustrated in the same
figure. For p$_{T}$ spectra, angular distributions and centrality
dependences, the subtraction procedure is done in narrow slices of the
respective variables (see [11-14] for details and error
discussion). The subtracted data for the $\eta$, $\omega$ and $\phi$
themselves are subject to the same further steps as the excess data.

The (extended) IMR part is shown in the right panel of
Fig.1~\cite{Arnaldi:2008er}. The use of the silicon vertex tracker has
allowed to disentangle prompt and offset dimuons resulting from
simultaneous semi-muonic decays of (correlated) D and $\bar{D}$
mesons. The results are perfectly consistent with no enhancement of
open charm, but rather match the level expected from upscaling the
NA50 results on charm in pA collisions. The excess observed in the
IMR, confirming the result previously found by NA50, is really prompt,
with an enhancement over Drell-Yan production of dimuons by a factor
of 2.4$\pm$0.08~\cite{Arnaldi:2008er}. The excess can now be isolated
in the same way as done in the LMR, by subtracting the expected known
sources, here Drell-Yan and open charm as plotted in Fig.1, from the
total data.

%%%%%% Fig.2
\begin{figure*}[t]
\includegraphics*[width=0.5\textwidth]{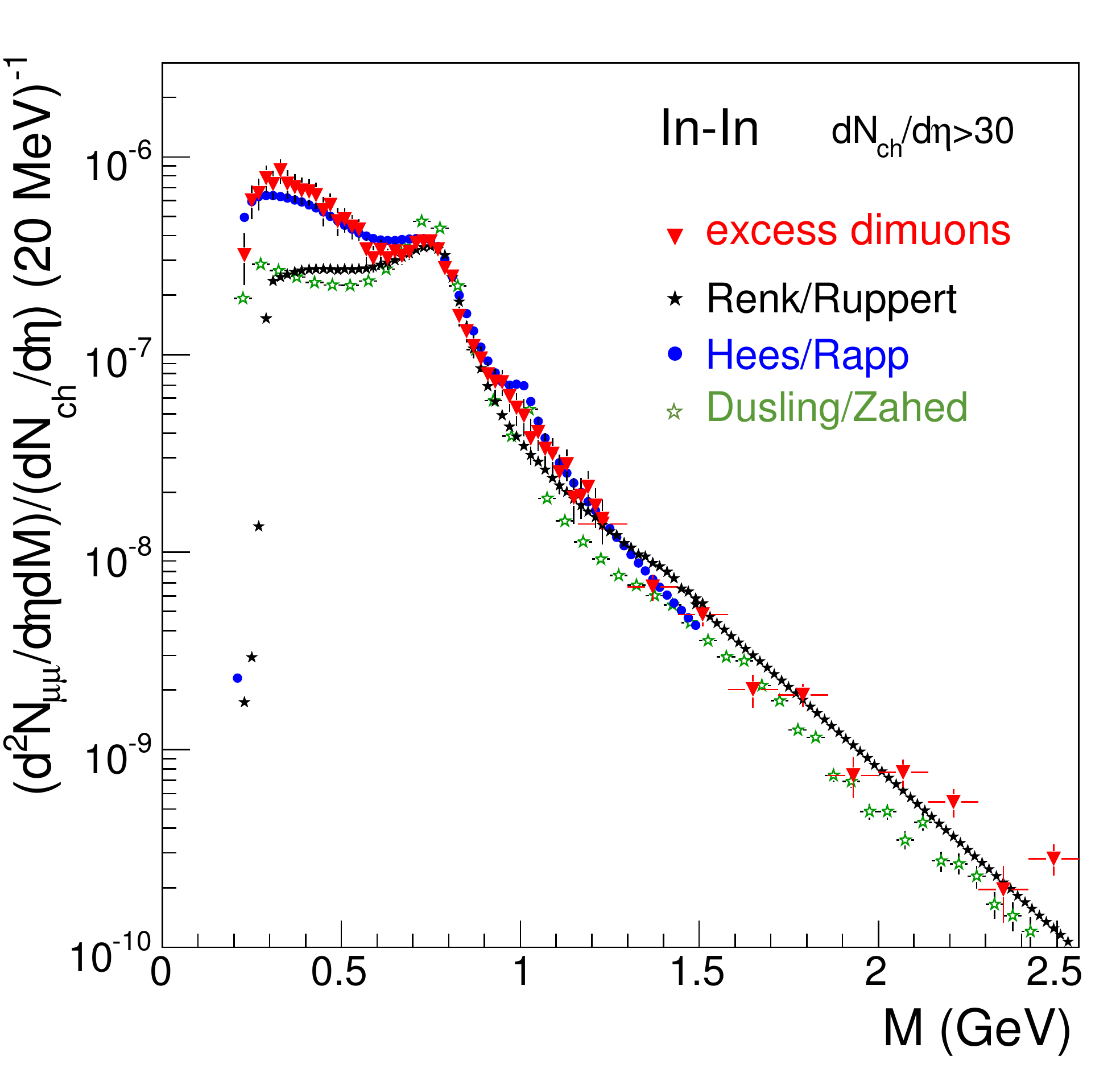}
\includegraphics*[width=0.5\textwidth]{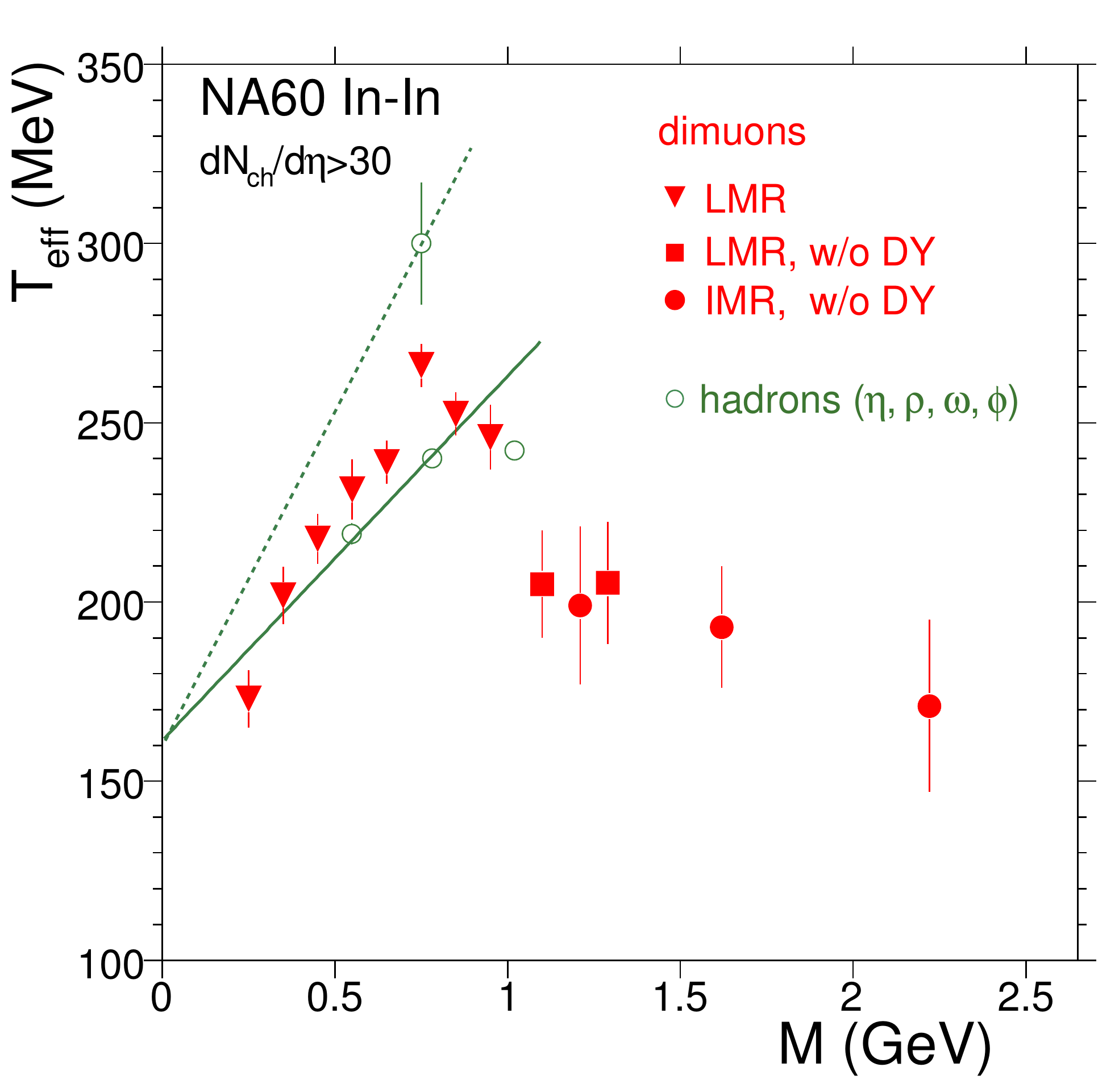}
\caption{Left Panel: Acceptance-corrected invariant mass spectrum of 
the excess dimuons, integrated over p$_{T}$, compared with three
different sets of thermal-model results in absolute terms. Right
panel: Inverse slope parameter T$_\mathrm{eff}$ of the
acceptance-corrected m$_{T}$ spectra vs. dimuon mass
(see~\cite{Arnaldi:2007ru} for a discussion of the statistical and
systematic errors). Hadron results are shown for comparison.}
\label{fig2}
\end{figure*}

In the last step, the data are corrected for the acceptance of the
NA60 apparatus and for the centrality-dependent pair reconstruction
efficiencies. The acceptance shows strong variations with mass and
p$_{T}$ in the low-mass/low-p$_{T}$
region~\cite{Damjanovic:2006bd,Damjanovic:2007qm}, but is understood
on the level of $<$10\%, based on a detailed study of the peripheral
data ~\cite{Damjanovic:2006bd} and a number of further
investigations. In principle, the correction requires a 5-dimensional
grid in M-p$_{T}$-y-cos$\theta$-cos$\phi$ space ($\theta$, $\phi$
being the angles describing the dilepton angular distributions). This
is neither realistic nor optimal. To avoid the large statistical
errors connected with low-acceptance bins, the correction is usually
done in 2-dimensional grids, using the measured distributions in the
other variables as an input (examples
M-p$_{T}$~\cite{Arnaldi:2007ru,Arnaldi:2008fw,Damjanovic:2007qm} or
y-p$_{T}$~\cite{Damjanovic:2007qm} or
$\theta$-$\phi$~\cite{Arnaldi:2008gp}). This clearly requires an
iteration procedure to become self-consistent, fortunately eased by an
often only small sensitivity. It is of great importance, however, to
treat the different sources separately, i.e. {\it after} the
subtraction procedure, due to differences in the distributions (open
charm, e.g., is completely different).

\section{THERMAL RADIATION}

The prime results on the excess data are summarized in Figs.2-4. The
pure data aspects will be discussed first, followed by a coherent
interpretation of all results in terms of thermal radiation further
below. The left panel of Fig.2 shows the inclusive invariant mass
spectrum of the excess dimuons for the complete range 0.2$<$M$<$2.5
GeV, with all known sources subtracted (except for the $\rho$),
integrated over p$_{T}$, corrected for experimental acceptance and
normalized absolutely to the charged-particle rapidity
density~\cite{Damjanovic:2009zz}. Compared to an earlier version of the 
figure~\cite{Arnaldi:2008er}, the errors have significantly decreased at low
masses, using a 1-dimensional acceptance correction in M based on the
measured correlated M-p$_{T}$ information~(\cite{Arnaldi:2008fw} and
Fig.3). In addition, the subtraction of the narrow resonances is now
based on an improved modeling of the experimental
resolution~\cite{Arnaldi:2009wr}, leading to a smoother spectral shape
in the region of the $\phi$. The left panel of Fig.3 shows the
associated m$_{T}$ spectra for the LMR part, where
m$_{T}$=(p$_{T}^{2}$+M$^{2}$)$^{1/2}$. The normalization is arbitrary,
allowing for an even spacing of the 10 spectra, but the absolute
normalization of their integrals can directly be taken from the mass
spectrum. In contrast to an earlier version of the
figure~\cite{Arnaldi:2007ru,Arnaldi:2008er}, the p$_{T}$ coverage has
mostly been extended to about 3 GeV. The m$_{T}$ spectra for the IMR
part are shown in the right panel of
Fig.3~\cite{Arnaldi:2008er,Arnaldi:2008fw}. All m$_{T}$ spectra are,
to a very high degree of accuracy~\cite{Damjanovic:2007qm}, pure
exponentials for (m$_{T}$-M)$\geq$0.2 GeV~\cite{Arnaldi:2007ru}.  The
complete information can therefore be condensed into one single
parameter for each spectrum (determined with a very high accuracy),
the inverse slope parameter T$_\mathrm{eff}$, obtained by fitting the
m$_{T}$ data with the usual ``m$_{T}$-scaling'' expression
$1/m_{T}dN/dm_{T}$$\sim$ $exp(-m_{T}/T_\mathrm{eff})$. The resulting
values of T$_\mathrm{eff}$ are plotted in the right panel of Fig.2
vs. dimuon mass. Above 1 GeV, the LMR data are corrected for DY to be
consistent with the (independent) IMR analysis. To complement the
information on M and p$_{T}$, Fig.4 finally shows the results from a
systematic study of the dimuon angular
distributions~\cite{Arnaldi:2008gp}, another first in the field. Using
the Collins-Soper reference frame (but the results are independent of
that choice), all structure function parameters $\lambda$, $\mu$ and
$\nu$ (related to the spin-density matrix of the virtual photon) are
found to be zero, and the projected distributions in |cos$\theta$| and
|$\phi$| are seen in Fig.4 to be uniform. This is a nontrivial result:
the annihilation of quarks or pions along the beam direction, e.g.,
would lead to $\lambda$=+1, $\mu$=$\nu$=0 (lowest-order DY) or
$\lambda$=-1, $\mu$=$\nu$=0, corresponding to transverse and
longitudinal polarization of the virtual photon, respectively.

%%%%%% Fig.3
\begin{figure*}[t]
\includegraphics*[width=0.48\textwidth]{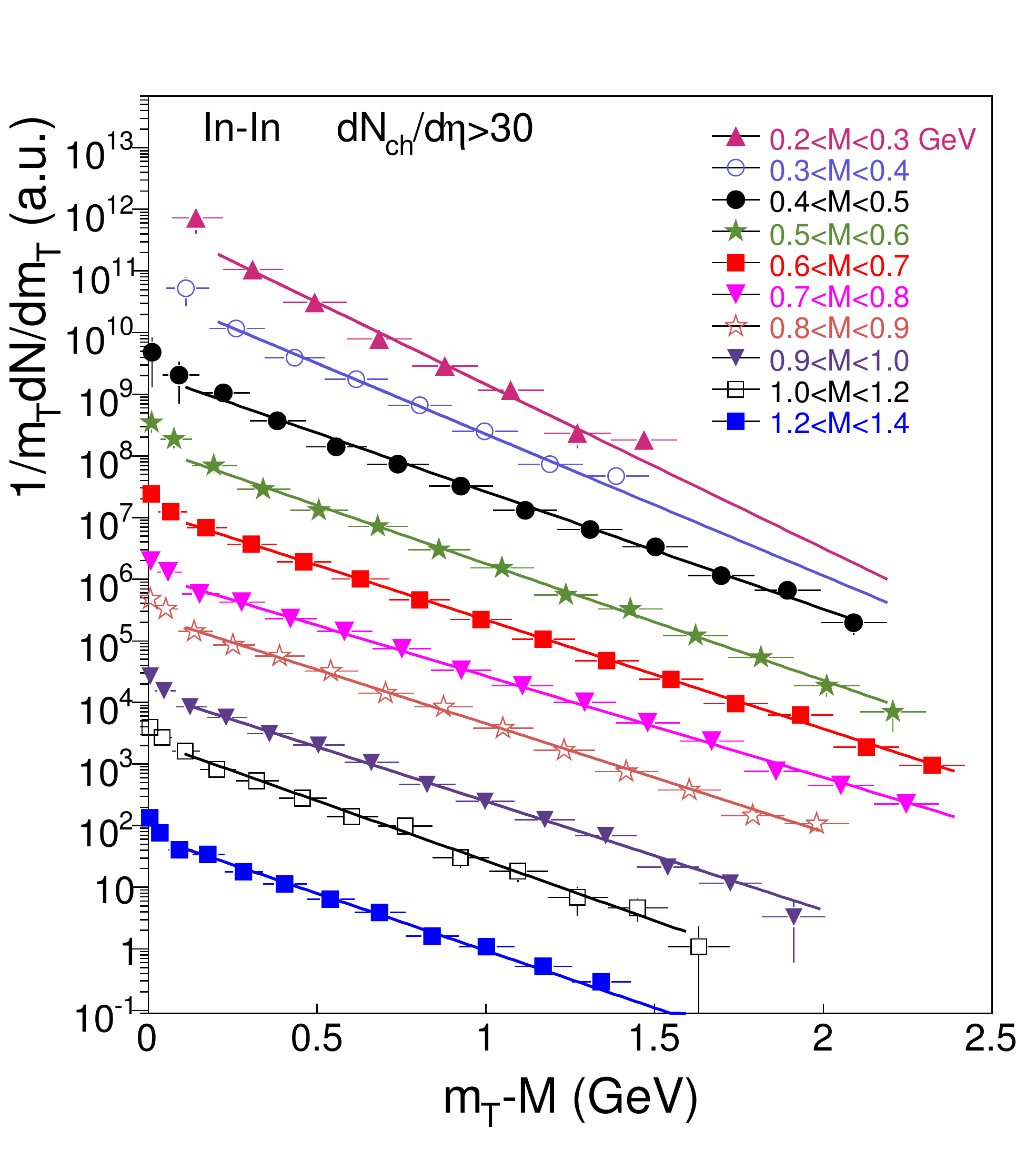}
\includegraphics*[width=0.48\textwidth]{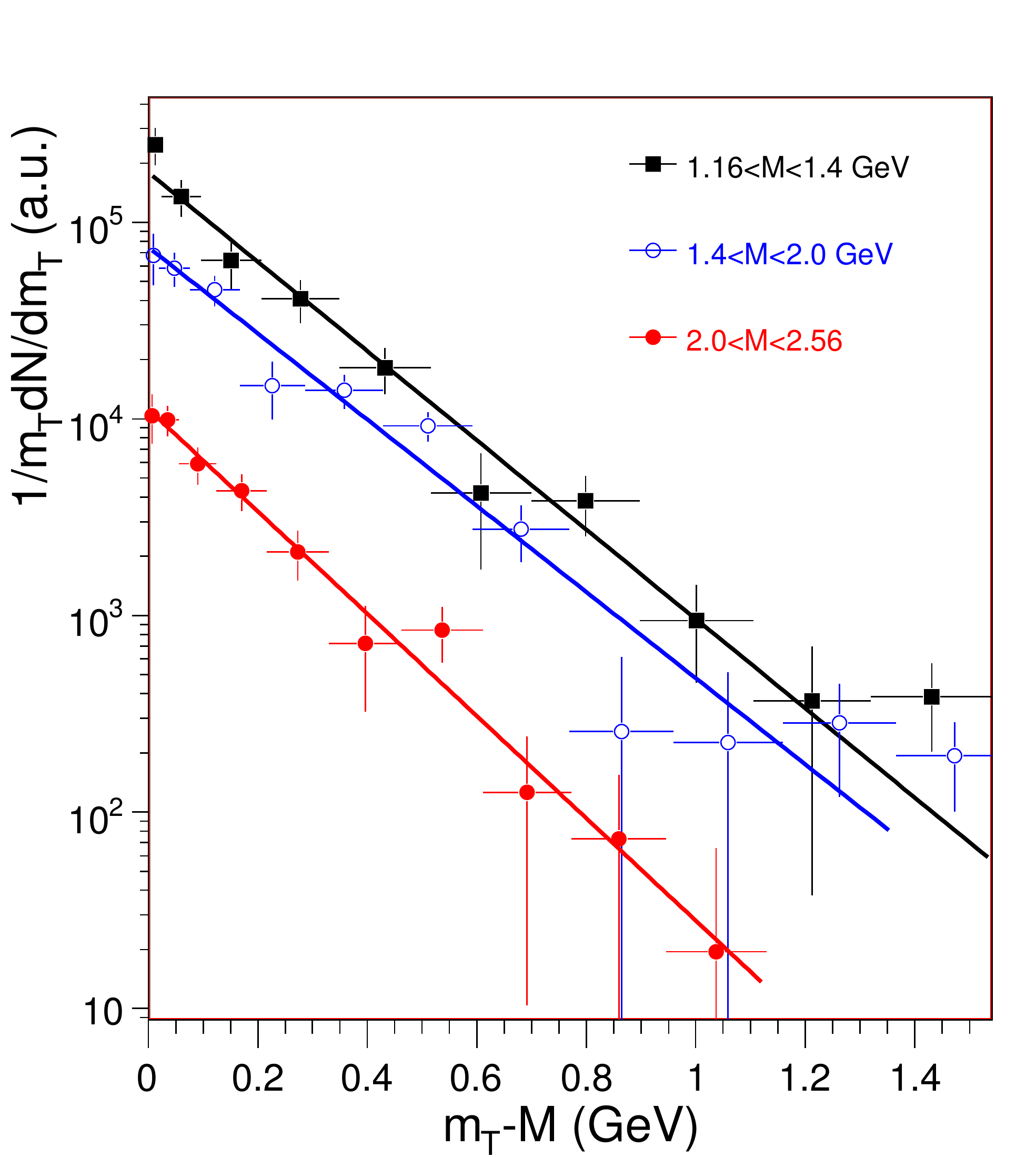}
\caption{Left Panel: Acceptance-corrected m$_{T}$ spectra of 
the excess dimuons for 10 mass windows in the LMR. Open charm is
subtracted throughout. Right panel: Acceptance-corrected m$_{T}$
spectra of the excess dimuons for 3 mass windows in the IMR.}
\label{fig3}
\end{figure*}

Without resorting to any detailed theoretical modeling, the data
themselves allow for a consistent global interpretation of the
observed excess dimuons in terms of thermal radiation from the
fireball. Three necessary prerequisites to justify the term are
clearly fulfilled: (i) a Planck-like (nearly exponential) shape of
the mass spectrum above 1 GeV, where the underlying spectral functions
are expected to be uniform (as in the black-body case), (ii) purely
exponential m$_{T}$ spectra, and (iii) the absence of any polarization
as expected for radiation from a randomized system.

%%%%%% Fig.4
\begin{figure*}[t]
\includegraphics*[width=0.5\textwidth]{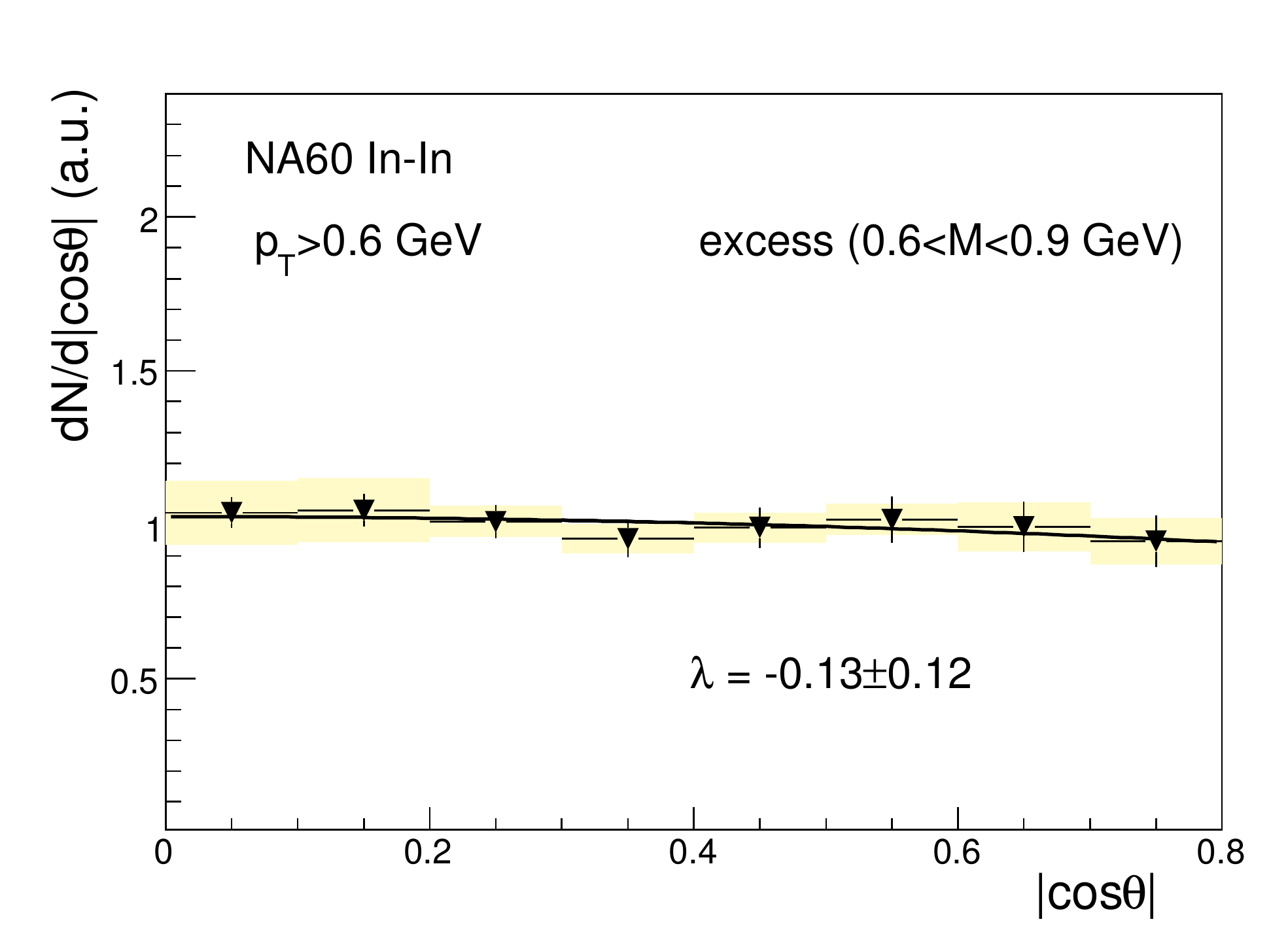}
\includegraphics*[width=0.5\textwidth]{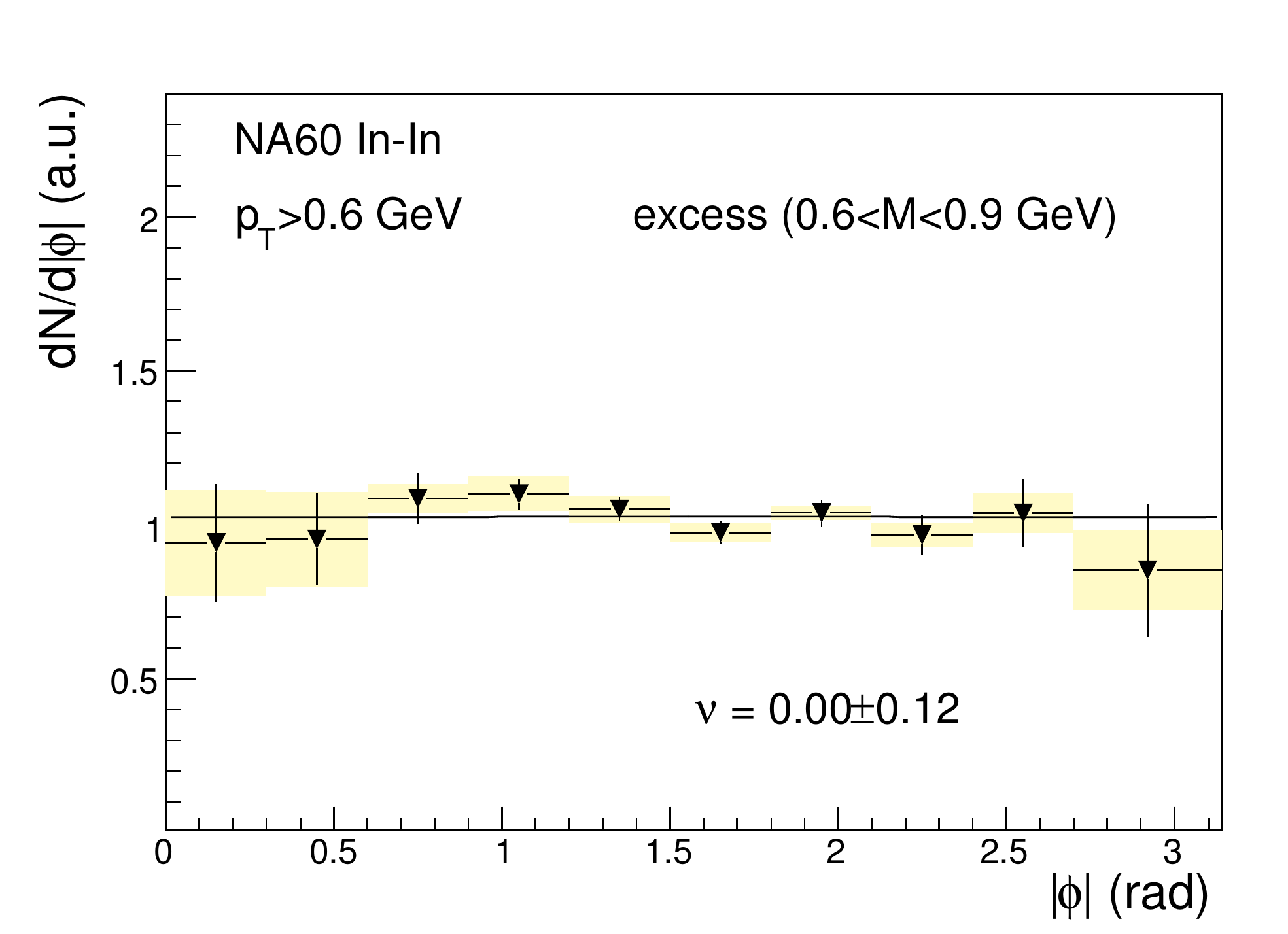}
\caption{Acceptance-corrected polar (left) and azimuthal (right) 
angular distributions of the excess dimuons. The Collins-Soper
   reference frame has been used here. Analogous results have been
   obtained for the mass window 0.4$<$M$<$0.6
   GeV~\cite{Arnaldi:2008gp}.} \label{fig4}
\end{figure*}

Turning next to a more detailed discussion, the large drop of
T$_\mathrm{eff}$ around masses of 1 GeV as seen in the right panel of
Fig.2 is a strong hint for a qualitative change of the radiation from
the LMR to the IMR. The mass region $<$1 GeV has traditionally been
associated with the process
$\pi^{+}\pi^{-}\rightarrow\rho\rightarrow\mu^{+}\mu^{-}$ as the
dominant dilepton source. Indeed, the mass spectrum shows a
considerable modulation, with clear signs for a (broadened) $\rho$. In
the same region, T$_\mathrm{eff}$ shows a monotonic rise with mass all
the way up to the nominal pole position of the $\rho$. This is a
strong indication for radial flow of a ``hadron-like'' dilepton
source, as confirmed by the genuine NA60 hadron data plotted in the
same figure. The parameter T$_\mathrm{eff}$ can roughly be described
by a superposition of a thermal and a flow part in the form
T$_\mathrm{eff}$$\sim$T+M$v$$^{2}$, where $v$ is the average
flow velocity. Maximal flow is reached for the point with
T$_\mathrm{eff}$$\sim$300 MeV, obtained by isolating the peak part in
the mass spectrum by a side-window subtraction
method~\cite{Arnaldi:2008fw} and interpreted as the freeze-out $\rho$
at the end of the fireball evolution, synonymous to the other
hadrons. The maximal flow then mirrors the expected maximal coupling
of the $\rho$ to pions, while all other hadrons freeze out earlier, in
quantitative support by a NA60 blast-wave
analysis~\cite{Arnaldi:2010zz}. Extrapolating the trend of
T$_\mathrm{eff}$ vs. M down to zero mass (and taking account of
relativistic corrections for M$<$p$_{T}$), the average temperature is
found to be about 130-140 MeV in this region, considerably below
T$_{c}$$\sim$170 MeV.

The interpretation of the mass region $>$1 GeV is quite
straightforward, eased by an argument not placed before. Assuming flat
spectral functions in the spirit of parton-hadron duality in the
dilepton rates, the momentum-integrated yield is given by $dN/dM\sim
M^{3/2}exp(-M/T)$, where T reflects some space-time average over the
system evolution. A fit of the mass spectrum with this expression over
the range 1.2$<$M$<$2.0 GeV gives T=205$\pm$12 MeV. Since M is by
construction a Lorentz-invariant, the mass spectrum is immune to any
motion of the emitting sources, unlike m$_{T}$ spectra or Planck's law
itself. The parameter T in the spectral shape of the mass spectrum is
therefore purely thermal. The fit value of about 200 MeV, considerably
above T$_{c}$$\sim$170 MeV, thus directly implies partonic emission
sources to dominate in the IMR. This is fully consistent with an
independent argument placed by NA60 from the
beginning~\cite{Arnaldi:2007ru,Arnaldi:2008fw}: the striking jump-like
drop in T$_\mathrm{eff}$ around 1 GeV visible in Fig.2 can never be
created by a continued hadronic scenario, but reflects early emission
with a possible connection to the soft point in the equation-of-state
as obtained from lattice QCD (most relevant at SPS energies). In the
partonic scenario, flow can hardly develop up to T$_{c}$. This is
reflected by the same average value of about 200 MeV seen in the
T$_\mathrm{eff}$ plot beyond 1 GeV, with no margin left for flow
within the errors of 10-20 MeV, not to speak about any rising mass
dependence in this region. These rather clear conclusions also end the
years-long discussion on hadron-parton duality, where duality in the
rates has been incorrectly taken over as duality in the yields,
ignoring the differences in the space-time integration for partons and
hadrons.

The three sets of thermal model results [21-23] which are also plotted
in the left panel of Fig.2 largely confirm the above observations. To
the extent that they roughly agree both with the shape and the
absolute magnitude of the data, they present a fourth strong argument
supporting the interpretation of the excess dimuons as thermal
radiation. In particular, the agreement of the Hees/Rapp
scenario~\cite{RH:2008lp} with the data in the region of the $\rho$
spectral function is really spectacular, as shown even more clearly in
a paper by Rapp presented at this Workshop~\cite{Rapp:2010sj}. In the
region above 1 GeV, all three models explicitly differentiate between
partonic and hadronic processes. In case
of~\cite{Renk:2008prc,Dusling:2007rh}, partons dominate, and the data
are fully described up to 2.5 GeV. The average temperature associated
with these two scenarios is 217 MeV (compared to the measured
205$\pm$12 MeV), with a small increase of the local values by about
10\% towards higher masses. Partons also dominate in a further
scenario not contained in Fig.2~\cite{Br:2009}. In case
of~\cite{RH:2008lp} however, hadrons dominate, with corresponding
lower values of T, but the predictions stop at 1.5 GeV, too low to
completely bear out the resulting conflict with the data in the region
up to 2.5 GeV.

\section{VECTOR MESON SPECTRAL FUNCTIONS}
%%\subsubsection{<A subsubsection>}

The mediation of thermal radiation in the LMR by the vector mesons
implies a convolution of the dilepton rate over the space-time
evolution of the fireball, where the rate is given by the photon
propagator, phase space factors and the vector meson spectral
functions~\cite{Rapp:1995zy}. In case of the narrow $\omega$ and
$\phi$, the convolution hardly affects the line shape, and these
mesons can therefore be isolated in a straightforward way. In case of
the broad $\rho$, however, in particular if further broadened in the
medium, the spectral shape is completely masked by the convolution
procedure as seen in Fig.2, where the dilepton yield even increases
further below the nominal pole of the $\rho$. Unfolding the final
result is plainly impossible. A realistic way to at least project out
the space-time averaged $\rho$ spectral function is the use of a
suitable correction function in 2-dimensional M-p$_{T}$ space. By a
strange coincidence of suitable conditions, the experimental
acceptance of NA60, being strongly reduced in the low-mass/low-p$_{T}$
region~\cite{Damjanovic:2006bd,Damjanovic:2007qm}, provides such a
correction function in an unexpectedly perfect way. The left panel of
Fig.5 proves the point~\cite{Damjanovic:2006bd}. Thermal radiation
with an underlying uniform spectral function, using a full (uncut)
p$_{T}$ spectrum with an inverse slope parameter T$_\mathrm{eff}$
representative for the average, is transformed after propagation
through the acceptance into again a uniform spectrum up to about 1
GeV, within 10\%. In other words, the steep rise of the thermal
spectrum due to the photon propagator and the Boltzmann factor is just
about compensated by the falling acceptance in this region. Variations
of the input p$_{T}$ spectrum within reasonable physics limits affect
the flatness of the output by at most 20\%.

The right panel of Fig.5 shows the excess data in the LMR as directly
measured, i.e. before correction for
acceptance~\cite{Arnaldi:2006jq,Arnaldi:2008fw}. This then can be
interpreted as the in-medium $\rho$ spectral function, averaged over
space-time and momenta. In order to compare to theory, a centrality
selection of 110$<$$dN_{ch}/d\eta$$<$170 has been applied, but the
data are very close to the inclusive data ($<dN_{ch}/d\eta>$=120) with
their much better statistics, contained in the middle panel of
Fig.1. The evolution of the $\rho$ spectral function with centrality
is discussed in~\cite{Damjanovic:2006bd,Arnaldi:2008fw}. A peaked
structure is always seen, broadening strongly with centrality, but
remaining essentially centered around the nominal pole position of the
$\rho$. The $rms$ of the distributions increases monotonically from
that of a free $\rho$ to almost the value of a uniform spectrum. The
total yield relative to the (estimated) cocktail $\rho$ increases by a
factor of 6-7, reflecting the ``$\rho$-clock''~\cite{Heinz:1991fn},
i.e. the number of $\rho$ generations created during the fireball
evolution.

Fig.5 also contains the two main theoretical scenarios developed
historically for the in-medium spectral properties of the $\rho$,
dropping mass~\cite{Brown:kk} and broadening~\cite{Rapp:1995zy},
evaluated here for the same fireball evolution~\cite{RH:2003}. The
model results are normalized to the data in the mass interval M$<$0.9
GeV, just to be independent of the fireball evolution. The unmodified
$\rho$ is clearly ruled out. The broadening scenario gets remarkably
close, while the dropping mass scenario completely fails. This ends a
decades-long controversy about the spectral properties of hadrons
close to the QCD phase boundary, kept active solely through
insufficient data quality. The connection to chiral symmetry
restoration remains an open theoretical issue, but the way chiral
partners ultimately mix is probably answered with ``complete
melting''.

%%%%%% Fig.5
\begin{figure*}[t]
\includegraphics*[width=0.51\textwidth]{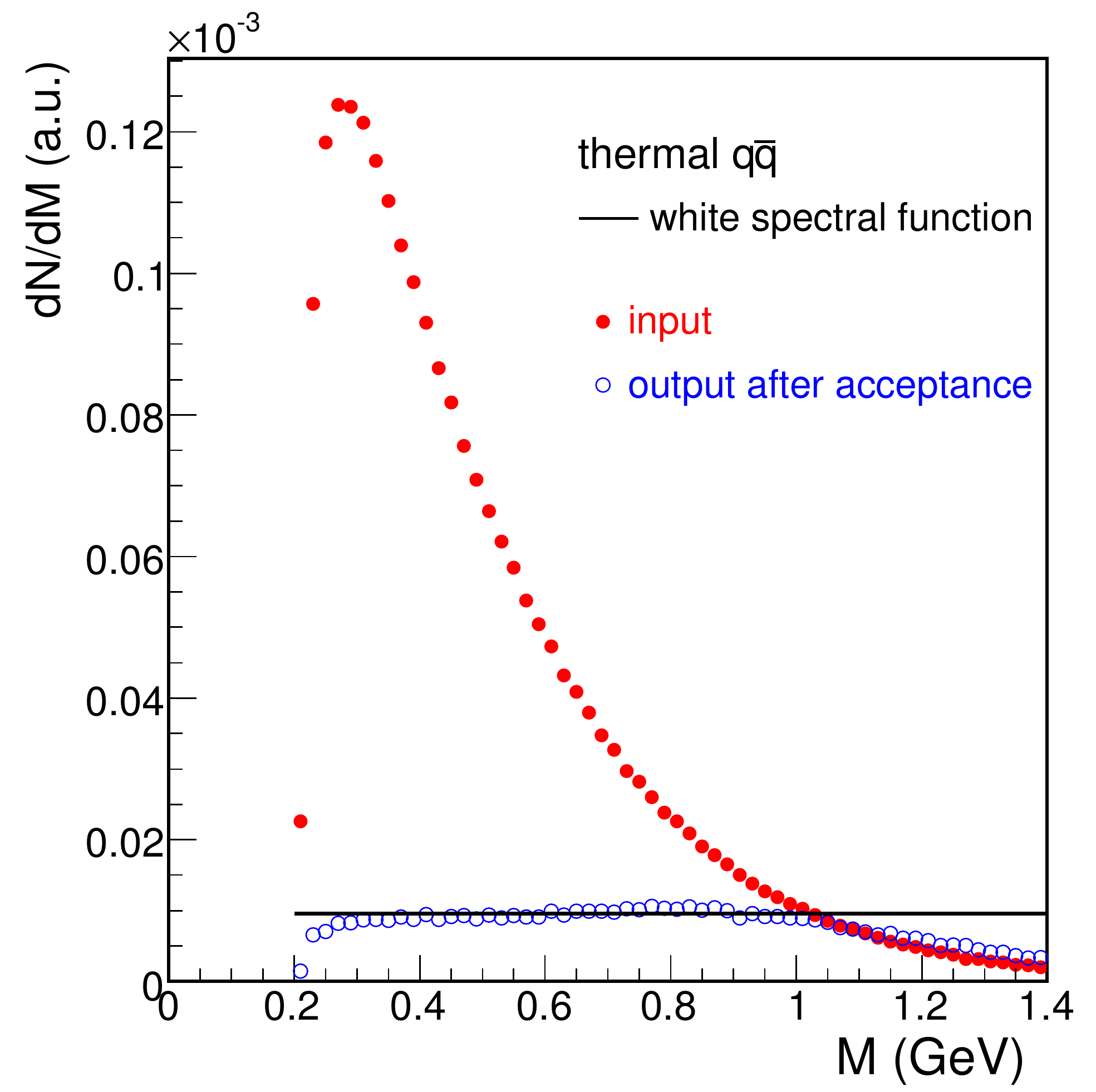}
\includegraphics*[width=0.5\textwidth]{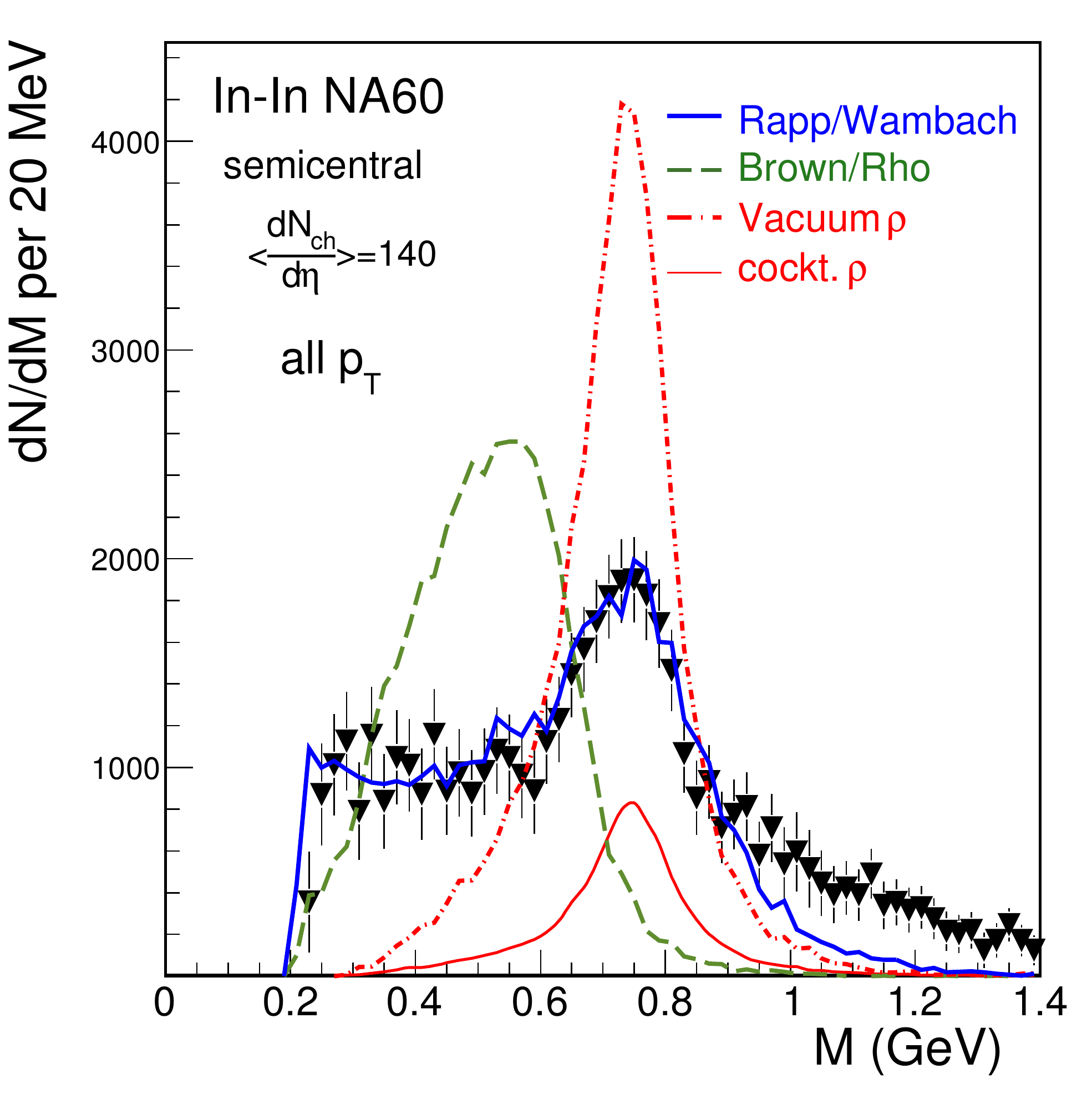}
\caption{Left panel: Propagation of thermal radiation based on a uniform 
spectral function through the NA60 acceptance. The resulting spectrum
is also uniform (see text). Right panel: Excess dimuons before
acceptance correction, reflecting the in-medium $\rho$ spectral
function averaged over space-time and momenta. The theoretical
predictions also shown are renormalized to the data for M$<$0.9
GeV. Beyond 0.9 GeV, other physical processes take over.}
\label{fig5}
\end{figure*}

In nuclear collisions, the longer-lived $\omega$ (23 fm) and $\phi$
(46 fm) have historically received much less attention than the
$\rho$, since most of their dilepton decays occur after thermal
freeze-out. The $\omega$ suffers, on top, from the strong masking by
the much more abundant (regenerated) $\rho$, leaving precision work
rather to cold nuclear matter experiments. NA60 has addressed the
$\omega$ in a way directly coupled to the cocktail subtraction
procedure. Due to the high mass resolution, the {\it disappearance} of
the yield at low p$_{T}$ out of the narrow $\omega$ peak in the
nominal pole position can sensitively be detected. The {\it
appearance} of the yield elsewhere in the mass spectrum, originating
from a mass shift or broadening or both is practically inaccessible,
due to the masking by the $\rho$. Evidence for the disappearance of
the $\omega$ has been found by the following
procedure~\cite{Arnaldi:2008fw}. The m$_{T}$ spectrum, measured to be
a pure exponential for (m$_{T}$-M)$\geq$0.6 GeV, is fit with the usual
m$_{T}$-scaling expression used before, defining a reference line.
The line is extrapolated to (m$_{T}$-M)=0, and the ratio
data/reference line is then used to monitor shape changes over the
complete range in m$_{T}$. The procedure is done separately for four
centrality bins. The results are shown in the left panel of
Fig.6~\cite{Arnaldi:2008fw}, absolutely normalized to the full phase
space ratio $\omega$/N$_{part}$. The effects of disappearance are
quite striking: (i) a suppression of the relative yield below the
reference line only occurs for p$_{T}$$\leq$1 GeV; (ii) there is a
very strong centrality dependence of the suppression, reaching down to
$\leq$0.5 of the reference line (the large errors reflect the
increasing masking by the $\rho$); (iii) the suppression effects are
much larger than expected for the spectral distortions due to the blue
shift from radial flow at low m$_{T}$; a simulation on the basis of
the blast wave analysis~\cite{Arnaldi:2010zz} shows at most 10\%
effects for central collisions. Thus, strong in-medium effects also
exist for the $\omega$, but it seems impossible to clarify their
nature within the NA60 frame.

%%%%%% Fig.6
\begin{figure*}[t]
\includegraphics*[width=0.5\textwidth]{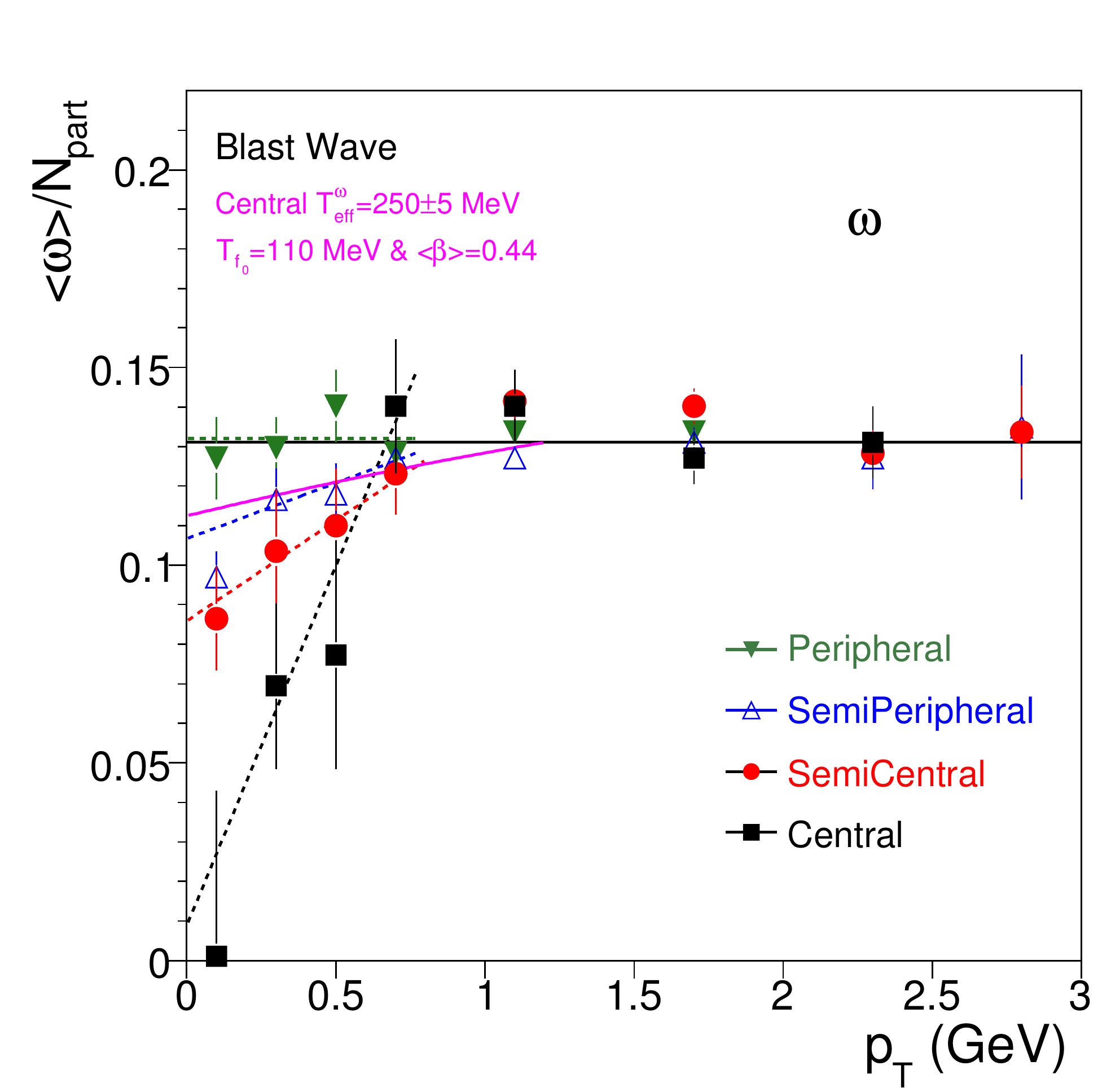}
\includegraphics*[width=0.5\textwidth]{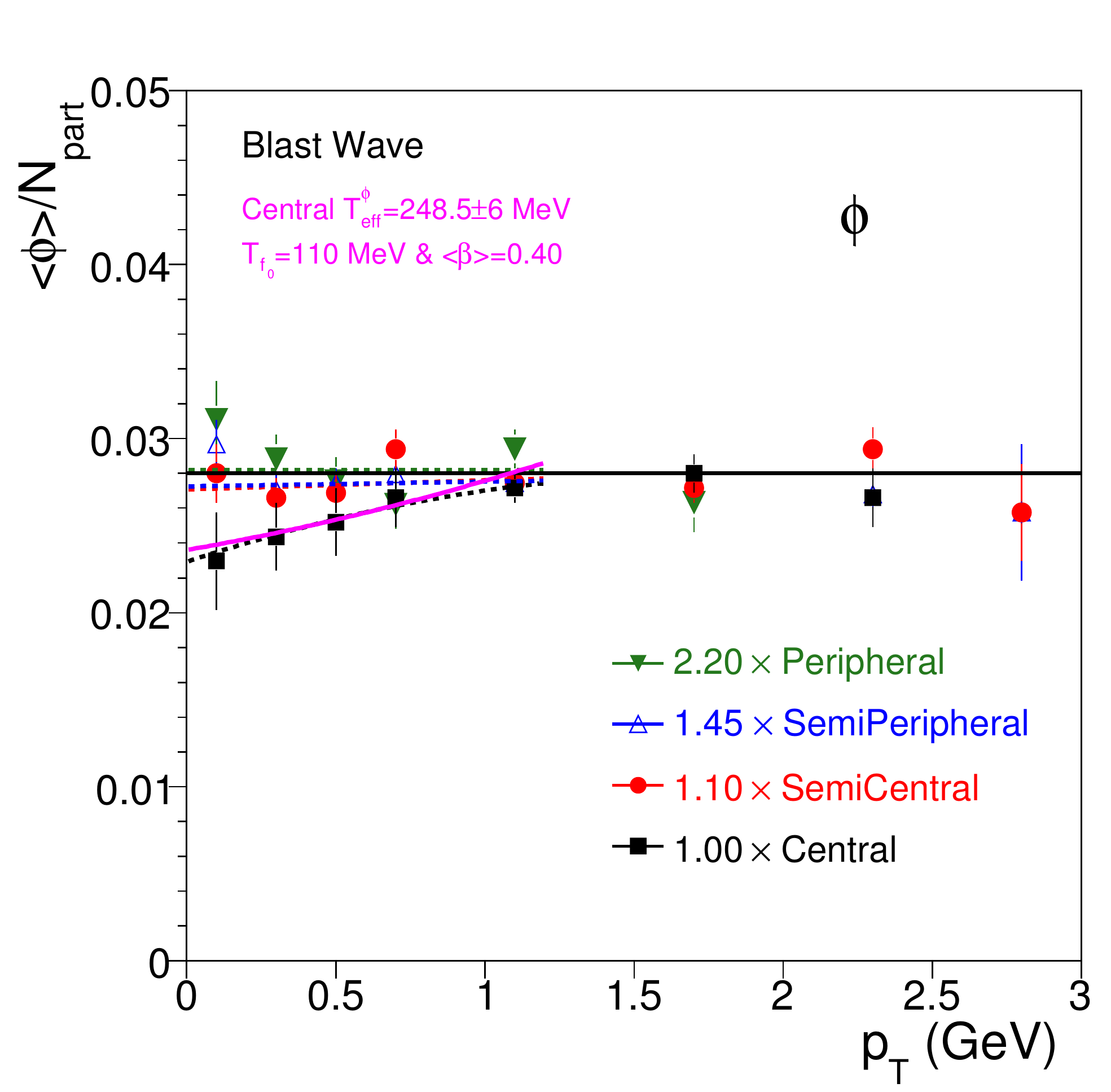}
\caption{p$_{T}$ dependence of the normalized yields of the $\omega$ 
(left) and $\phi$ (right) for 4 different centralities. The solid
lines for p$_{T}$$\leq$1 GeV below the reference lines show the
results from blast wave fits for central collisions. The dotted lines
are only meant to guide the eye. The errors are purely statistical;
the systematic errors are negligibly small compared to the statistical
ones.}  \label{fig6}
\end{figure*}

The results from the same procedure applied to the $\phi$ are shown in
the right panel of Fig.6. No effects are seen beyond the limit set by
the blast wave analysis. A precision analysis of the position and line
shape of the $\phi$, the most isolated and therefore ``easiest''
hadron case in the NA60 data, has also been
performed~\cite{Arnaldi:2009wr}. The mass and width are found to be
compatible with the PDG values at any centrality and at any p$_{T}$;
no evidence for in-medium modifications is observed.

\section{CONCLUSIONS}

The precision of the NA60 data has set new standards in this
field. This has allowed to consistently interpret the results in terms
of thermal radiation, including identification of the two major
sources. Theoretical modeling has matched this precision for the
in-medium $\rho$ spectral function, but not yet for a description of
the dynamics. It is hoped that the strong constraints placed by these
aspects of the data will contribute towards convergence, including the
(theoretically) still controversial dominance of partons in the IMR
already at SPS energies.

%%%%%%%%%%%%%%%%%%%%%%%%%%%%%%%%%%%%%%%%%%%%%%%%
%% BACKMATTER
%%%%%%%%%%%%%%%%%%%%%%%%%%%%%%%%%%%%%%%%%%%%%%%%

%% \begin{theacknowledgments}
%% \end{theacknowledgments}

%%%%%%%%%%%%%%%%%%%%%%%%%%%%%%%%%%%%%%%%%%%
%% Just a reminder that you may have to run bibtex
%% All of it up to \end{document} can be removed
%% if you don't like the warning.
%%%%%%%%%%%%%%%%%%%%%%%%%%%%%%%%%%%%%%%%%%%
%%%%\IfFileExists{\jobname.bbl}{}
%%%% {\typeout{}
%%%%  \typeout{******************************************}
%%%%  \typeout{** Please run "bibtex \jobname" to optain}
%%%%  \typeout{** the bibliography and then re-run LaTeX}
%%%%  \typeout{** twice to fix the references!}
%%%%  \typeout{******************************************}
%%%%  \typeout{}
%%%%  }

\end{document}